\documentclass[journal=apchd5,communication=article]{achemso}

\usepackage[version=3]{mhchem} 
\usepackage[T1]{fontenc}       
\usepackage{caption}
\usepackage{float}
\usepackage{geometry}
\usepackage{natbib}
\usepackage{setspace}
\usepackage{xkeyval}

\author{Ali Eshaghian Dorche}
\author{Sajjad Abdollahramezani}
\author{Hossein Taheri}
\author{Ali Asghar Eftekhar}
\author{Ali Adibi}
\email {ali.adibi@ece.gatech.edu}
\affiliation[Georgia Institute of Technology]
{School of Electrical and Computer Engineering, Georgia Institute of Technology, 778 Atlantic Drive NW, Atlanta, GA 30332, USA}

\title[An \textsf{achemso} demo]
 {Extending chip-based Kerr-comb to visible spectrum by dispersive wave engineering}

\abbreviations{}
\keywords{Dispersion engineering, Cherenkov radiation, coupled microresonators, silicon nitride, Kerr frequency comb, optical solitons}

\begin{document}
	\begin{abstract}
		Anomalous group velocity dispersion is a key parameter for generating bright solitons, and thus wideband Kerr frequency combs. Extension of frequency combs to visible wavelength in conventional photonic materials and structures has been a major challenge due to strong normal material dispersion at the relevant frequencies. Extension of frequency combs toward the normal dispersion region is possible via dispersive waves through soliton-induced Cherenkov radiation. However, this potentially powerful technique has not been used for extending frequency combs to the visible spectrum. In this paper, we demonstrate a new microresonator structure formed by an over-etched silicon nitride waveguide that enables the use of soliton-induced Cherenkov radiation to extend the bandwidth of the Kerr-combs. Furthermore, we show that by careful dispersion engineering in a coupled microring resonator structure we can optimize the properties (e.g., wavelength, and amplitude) of the generated dispersive wave to further extend the Kerr frequency combs to the visible spectrum while increasing the total Kerr-comb bandwidth as well.
	\end{abstract}
	
	\section{Introduction}
	Optical Frequency combs, i.e., equidistant narrow-linewidth pulses in the frequency domain with their ability to generate ultrashort pulses in time domain, are of great recent interest for numerous applications including precise time and distance measurement \cite{li2008laser,diddams2000direct,udem2002optical}, molecular fingerprint detection/sensing \cite{diddams2007molecular,thorpe2007cavity,schliesser2014mid}, astronomical ranging and spectroscopy \cite{yi2016demonstration,zou2016broadband,okawachi2011octave}, and optical communications \cite{pfeifle2014coherent,newbury2011searching}. Frequency combs can be generated either through mode locking in lasers \cite{udem1999accurate,sutter1999semiconductor} or through nonlinear Kerr effect in resonators \cite{del2007optical,kippenberg2011microresonator,del2011octave,foster2011silicon}. Kerr-combs are generated by virtue of third-order nonlinearity of optical materials at the presence of strong optical field enhancement in high finesse microresonators, which results in cascaded four-wave-mixing (FWM) and broad comb signal generation. Kerr-comb generation in microresonators has significant advantages over that using mode-locked laser including the possibility of on-chip integration in CMOS-compatible platforms and the possibility of achieving broadband comb signals. However, the generation of wideband Kerr-combs requires precise engineering of the microresonator dispersion over the desired bandwidth.
	
	Among different materials for generation of Kerr-combs, silicon nitride (SiN) is of great interest due to its very low optical loss in a large bandwidth (covering visible, near-infrared (NIR), and a good part of infrared (IR) wavelengths), ease of fabrication, and compatibility with CMOS fabrication processes. Recent reports show the feasibility of generation of Kerr-combs in SiN microresonators in IR \cite{levy2010cmos,pfeifle2014coherent,kippenberg2011microresonator,okawachi2011octave,riemensberger2012dispersion}. Several approaches have been proposed for engineering the dispersion of SiN ring/nanowire resonators based on engineering waveguide geometrical parameters (i.e., resonator height and width) \cite{moss2013new,boggio2014dispersion}, deposition of thin conformal dielectric layers with different refractive indices using atomic layer deposition (ALD) \cite{riemensberger2012dispersion}, taking advantage of mode crossing in coupled waveguide/slot waveguide structures (that are bent to form ring resonators) \cite{zhang2010flattened,bao2015increased,jafari2016parameter,ryu2012effect,de2011dispersion,zhu2012ultrabroadband,zhang2012silicon,zhang2013generation}, and introducing the layered waveguide configuration \cite{yang2016broadband,ainslie1986review,boggio2014dispersion}. More recently dispersive wave generation (also called soliton-induced Cherenkov radiation) was proposed as another approach to enhance bandwidth \cite{brasch2016photonic,akhmediev2016cherenkov}. Despite impressing recent progress in the generation of Kerr-combs, all reported results were in the IR spectrum, and the generation of Kerr-combs in visible wavelengths is still a major challenge.
	
	In this paper, we demonstrate a well-engineered SiN waveguide profile with its anomalous dispersion shifted to shorter wavelengths to enable the generation of frequency combs with spectral coverage from visible (around 700 nm) to NIR (around 1200 nm) wavelengths. We further substantiate the enhancement of comb bandwidth and power transfer efficiency toward visible wavelengths through the Cherenkov radiation process and demonstrate a coupled-racetrack resonator architecture to engineer/adjust the Cherenkov radiation wavelength to enable fine-tuning of the generated visible combs.
	
	\section {Dispersion engineering in over-etched waveguide}
\begin{figure*} 
		\centering
		\includegraphics[trim=0cm 2cm 0cm 0cm,width=16cm,clip]{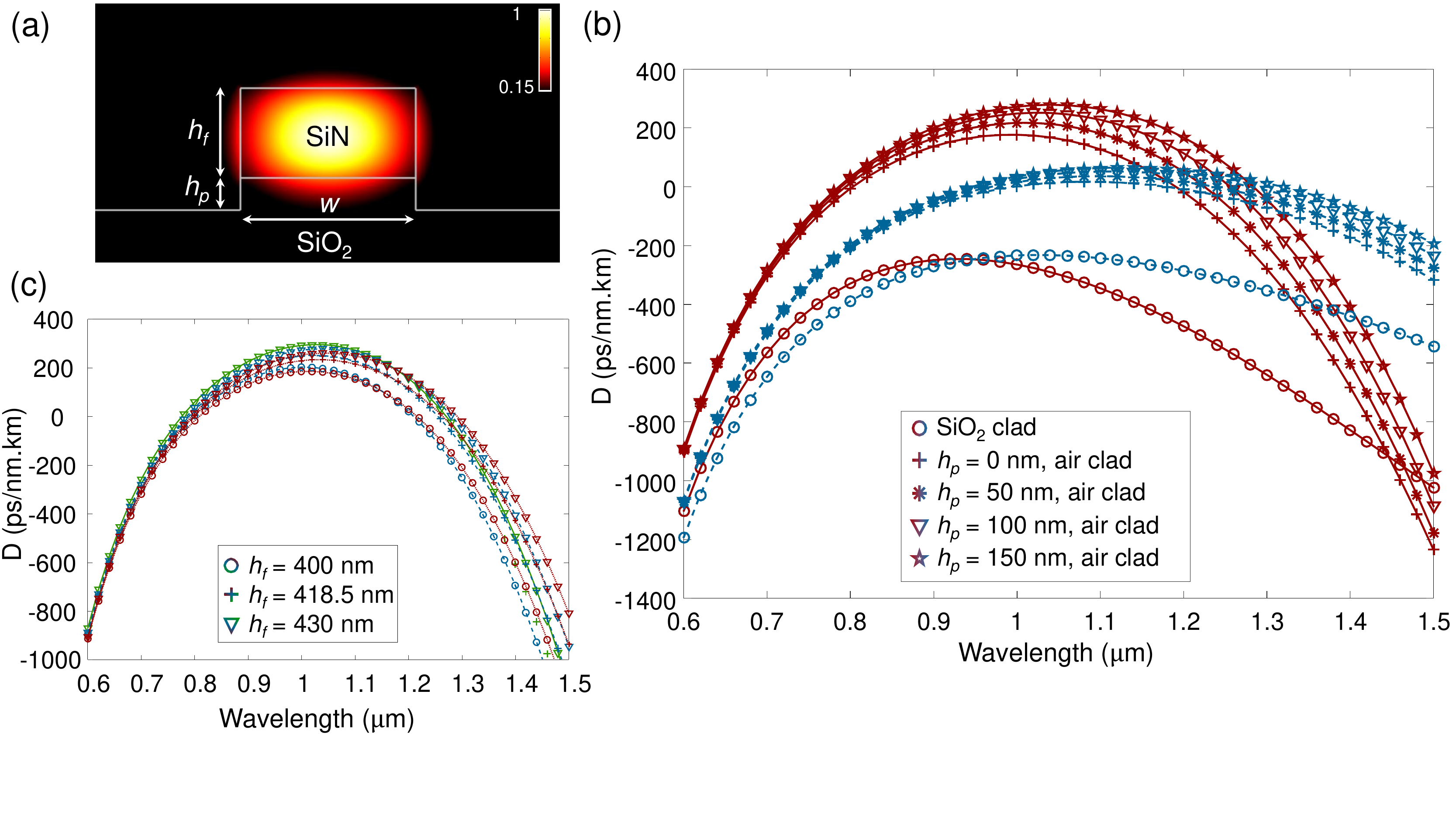}\\
		\captionsetup{justification=justified}
		\caption{Dispersion engineering of a thin over-etched SiN waveguide: a) mode profile and cross section of the over-etched SiN waveguide on $\mathrm{SiO}_{2}$ substrate with air cladding. b) Effect of width ($w$) and over-etching depth ($h_p$) on group velocity dispersion (GVD) of the waveguide in (a). Red and blue curves show the effect of $\mathrm{SiO}_{2}$ pedestal on GVD parameter for a SiN film with $w = 800$ nm and $w = 1100$ nm, respectively. In both cases, symmetric cladding ($\mathrm{SiO}_{2}$ clad with no over-etching), and different over-etching depths from 0 to 150 nm in steps of 50 nm are studied. c) GVD parameter for the proposed waveguide configuration in (a) with different widths and heights of SiN. Green, blue, and red curves are related to $w = 780$ nm, $w = 800$ nm, and $w = 820$ nm waveguide width. Three SiN thicknesses of $h_f = 400$ nm, $h_f = 417$ nm, and $h_f = 430$ nm are studied.}\label{fig1}
	\end{figure*}

	The proposed waveguide structure is shown in Fig. \ref{fig1}(a). The structure is composed of a 417 nm SiN film over silicon dioxide ($\mathrm{SiO}_{2}$). The ridge waveguide is formed by completely etching SiN as well as a portion of $\mathrm{SiO}_{2}$ to reduce the leakage of optical field. The over-etching of $\mathrm{SiO}_{2}$ (shown by the height $h_p$ in Fig. \ref{fig1}(a)) provides another degree of control (in addition to the waveguide width $w$ and SiN height $h_f$, in Fig. \ref{fig1}(a)) for engineering of the dispersion of the guided mode. This concept is similar to those used in multi-index fibers \cite{ainslie1986review} and multi-edge ring resonators \cite{boggio2014dispersion,yang2016broadband} for dispersion engineering. Most of the previous designs to achieve comb generation in SiN waveguides (at longer wavelengths) are based on thick SiN films (e.g., 800 nm), which makes fabrication process more challenging due to the effect of the built-up stress between the SiN film and the $\mathrm{SiO}_{2}$ substrate. Our proposed configuration in Fig. \ref{fig1}(a) blue shifts anomalous dispersion to shorter wavelengths while avoiding requirement for thick SiN. Figure \ref{fig1}(b) shows the group velocity dispersion spectra of the proposed waveguide structure for different over-etching depths ($h_p$). It is clear from Fig. \ref{fig1}(b) that the waveguide dispersion depicts stronger anomalous dispersion (larger $|D|$) over a wider bandwidth by increasing the over-etched depth $h_p$. The results in Fig. \ref{fig1}(b) were calculated using the finite-element method (FEM) package in the Comsol Multiphysics environment with modal analysis module.
	
	Figure \ref{fig1}(b) suggests that over-etching the waveguide into $\mathrm{SiO}_{2}$ increases anomalous dispersion bandwidth. In addition, it shows that the spectral range of anomalous dispersion extends with reducing the waveguide width. These observations can be explained by noting that the over-etching of the SiN waveguide into the $\mathrm{SiO}_{2}$ substrate increases the distance between the center of the field profile of the waveguide mode and the unetched $\mathrm{SiO}_{2}$ substrate (see Fig. \ref{fig1}(a)) through the addition of the barrier caused by the $\mathrm{SiO}_{2}$ pedestal with height $h_p$. It is known that the use of an asymmetric waveguide results in a non-zero cut-off frequency for the fundamental guided mode. In addition, the reduction of the waveguide width results in a blue-shift in this cut-off frequency, and thus a blue-shift in the zero dispersion wavelength caused by stronger anomalous waveguide dispersion (which compensates the normal material dispersion) \cite{boggio2014dispersion}. The combination of these effects results in the extension of the anomalous dispersion spectrum of the waveguide in Fig. \ref{fig1}(a) toward shorter wavelengths by using a narrower waveguide (smaller $w$ in Fig. \ref{fig1}(a)) with stronger over-etching (larger $h_p$ in Fig. \ref{fig1}(a)).
	
	It is obvious from Fig. \ref{fig1}(b) that a SiN waveguide with $w = 800$ nm, $h_t = 417$ nm, and $h_p = 150$ nm demonstrates anomalous dispersion around 900 nm which is desirable as the pumping wavelength in comb generation. For the rest of this paper, we use this waveguide configuration unless otherwise stated. To observe the sensitivity of the dispersion properties of the waveguide to variation of its geometrical parameters (e.g., caused by fabrication imperfection), Fig. \ref{fig1}(c) illustrates the variation of the dispersion parameter of the fundamental TE guided mode of the proposed waveguide when its width ($w$) and SiN height ($h_f$) vary in a large range ($\pm20$ nm and $\pm 15$ nm, respectively). As seen in Fig. \ref{fig1}(c), the waveguide dispersion does not change significantly, especially in the short wavelength region (e.g., 800-1000 nm) despite the relatively large change in the waveguide parameters. The results in Fig. \ref{fig1}(c) show a negligible $0.6 \%$ variation of the dispersion at $\lambda = 900$ nm wavelength per $1$ nm change in $w$ and $0.8 \%$ variation of dispersion at $\lambda = 900$ nm wavelength per $1$ nm change in $h_f$.
	
	\section{Kerr-comb generation and dispersive wave engineering}
	To study Kerr-comb generation, we consider a microring resonator with radius $40$ ${\mu}m$ formed based on the waveguide in Fig. \ref{fig1}(a). To calculate the nonlinear signal generation and evolution in the microring resonator under CW laser pumping, we solve the normalized Lugiato-Lefever equation (LLE) \cite{chembo2013spatiotemporal,godey2014stability} using the split-step Fourier method \cite{yunakovsky2006split,weideman1986split} and consider exact dispersion of the microring resonator instead of approximating dispersion with limited polynomial terms. In addition, 620 azimuthal modes of the microring resonator around the pumping wavelength are considered to ensure accuracy of the results. To ensure wideband Kerr-comb generation, the input pump power is set at a value higher than the threshold power ($P_{th}$) required for parametric frequency conversion given by \cite{matsko2005optical,chembo2010spectrum,herr2012universal,chembo2016quantum}
	\begin{equation}
	\label{eq:pth}
	P_{th}=\dfrac{{\kappa^{2}}{n_{0}^2}{V_\mathrm{eff}}}{8{\eta}c{\omega_{0}}n_{2}},
	\end{equation}
	where $\kappa$ denotes the cavity decay rate; $\eta$ is the coupling strength (which is equals to $\frac{1}{2}$ for critical coupling); $n_{0}$ and $n_{2}$ represent linear and nonlinear refractive indices of SiN at pumping wavelength, respectively; $V_{eff}$ is the effective nonlinear cavity volume; $c$ is the speed of light in free-space; and $\omega_{0}$ is the cold cavity resonance angular frequency.
	
	Considering critical coupling, loaded quality factor of $10^5$ (which is easily achievable for the selected microring resonator)  and pumping at $\lambda = 900$ nm, the threshold power is calculated to be $P_{th} = 0.15$ W. 
	The generalized LLE considering higher-order dispersion terms is \cite{chembo2013spatiotemporal,coen2013modeling}
	\begin{equation}
	\label{eq:lle}
	\dfrac{\mathrm{d}\psi}{\mathrm{d}\tau}=\\
	-(1+i\alpha)\psi+(\sum_{n=2}\dfrac{(-i)^{n-1}}{n!}(\dfrac{-2D_{n}}{\Delta{\omega}})\dfrac{\mathrm{d^{n}}\psi}{\mathrm{d}\theta^{n}})+i{\left|\psi\right|}^{2}\psi+F\\
	\end{equation}
	
	where $\psi$ is the normalized intracavity electric-field amplitude; $\alpha=-2(\Omega_{0}-\omega_{0})/\Delta\omega$ is normalized detuning, $\Omega_{0}$ is the angular frequency of the pumping laser, $F = (2g_0/{\Delta\omega_{0}})^{1/2}F_{0}^{*}$ is the normalized input power with $F_0$ as the amplitude of external excitation; and $D_{n}$ is the $n^{th}$ order dispersion parameter in the polynomial (Taylor) expansion of the microring resonator dispersion around the pumping frequency; furthermore, $\theta \in [-\pi, \pi]$ is the azimuthal angle, $\tau = \Delta\omega_{0}t/2$ is rescaled time, and $\Delta\omega_{tot}$ represents the linewidth of the resonance mode. The input pumping power for the simulations is set at 451.8 mW (753 mW) corresponding to ${F}^2 = 3(5)$.
	
\begin{figure*} 
		\centering
		\includegraphics[trim=0cm 1cm 0cm 0cm,width=16cm,clip]{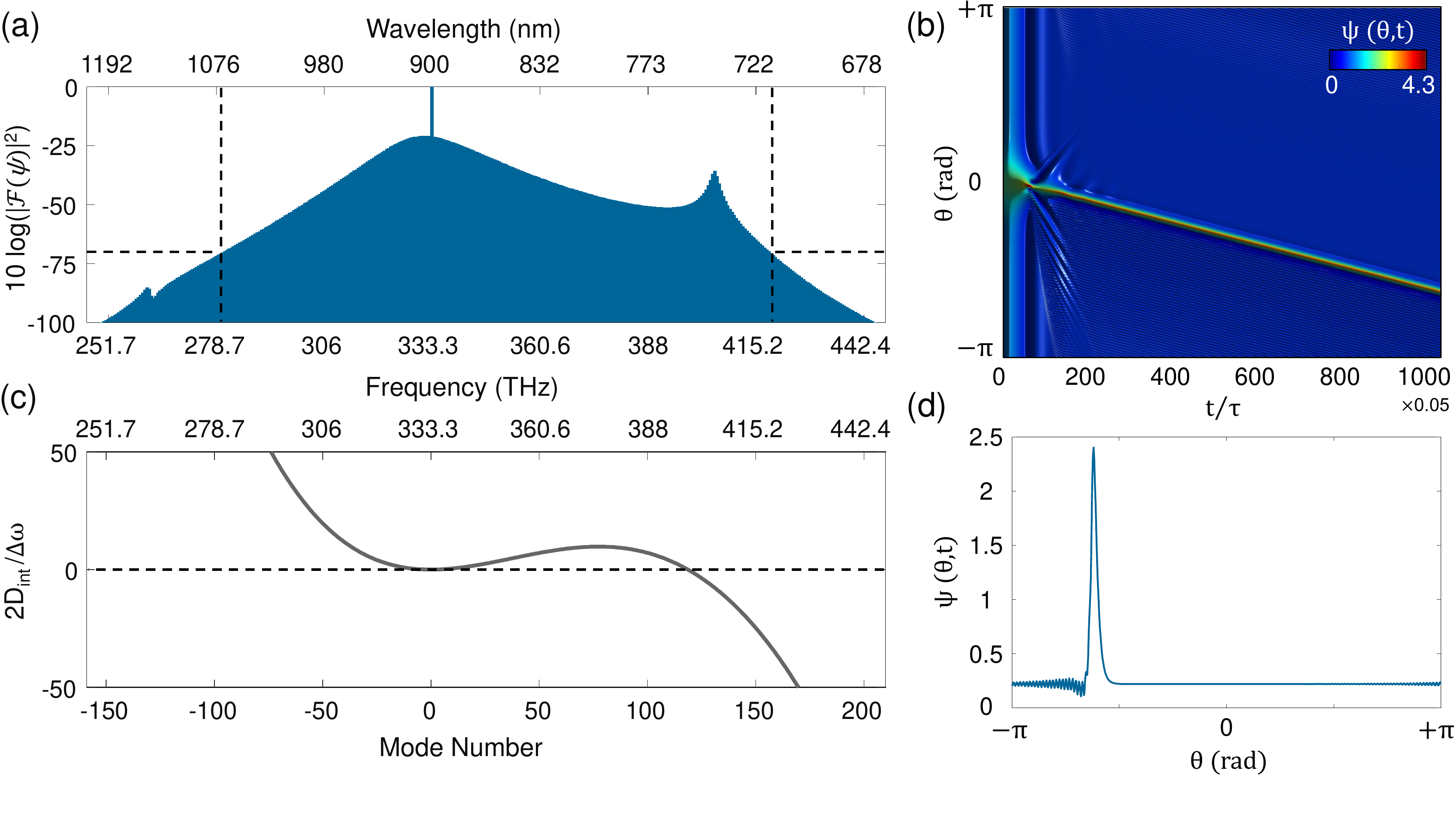}\\
		\captionsetup{justification=justified}
		\caption{Single soliton and soliton-induced Cherenkov radiation in a SiN microring resonator: a) Kerr-comb spectrum of the 40 $\mu$m-radius microring resonator formed using the proposed waveguide geometry. The peak in intensity at $\lambda =$ 741 nm (deep red color wavelength) arose as a result of Cherenkov radiation. The pump wavelength is 900 nm, normalized detuning is $\alpha = 3 $, and normalized power is $F^2 = 3$. The spectral range of the generated Kerr-comb at $-70$ dB window (i.e., $715$ nm - $1070$ nm) is shown by dashed lines. b) Time evolution of the signal in the microring resonator, which represents the formation of a soliton. c) Normalized integrated dispersion ($2D_{int}/{\Delta\omega}$) in the 40 $\mu$m-radius microring resonator formed using the proposed waveguide geometry in Fig. \ref{fig1}(a). d) Soliton amplitude formed in the resonator and its tail, which indicates Cherenkov radiation. }\label{fig2}
	\end{figure*}

	Figure \ref{fig2} depicts the result of LLE simulations for the $40\ \mu$m radius microring resonator described earlier with a normalized detuning of $\alpha = 3$ and normalized power of $F^{2} = 3$ ($451.8$ mW pump power). Figure \ref{fig2}(a) shows the normalized power spectrum in the resonator (pumping wavelength $900$ nm). Figure \ref{fig2}(a) clearly shows that the designed structure with the given parameters enables the formation of a single soliton with soliton-induced Cherenkov radiation (at $741$ nm); and the generation of an extended frequency comb with spectrum from NIR to the visible wavelength range.
	Optical solitons emit dispersive waves via the procedure entitled soliton Cherenkov radiation when perturbed by higher order dispersion  \cite{brasch2016photonic,akhmediev2016cherenkov,akhmediev1995cherenkov}. In this procedure, soliton transfers energy to the dispersive wave at the resonance wavelength that is the wavelength at which Kerr-comb spectrum experiences another peak (i.e., the peak at $741$ nm in Fig. \ref{fig2}(a)). To calculate the Cherenkov radiation wavelength, linear dispersion relation corresponding to Eq. (\ref{eq:lle}) is calculated approximately by substituting $\psi = \mathrm{exp}(ik_{lin}z+im\theta)$ into the normalized LLE and neglecting the nonlinear terms \cite{akhmediev1995cherenkov}. Soliton with wavenumber $k_{sol}=A^{2}/2$ will be at resonance with the dispersive wave (i.e., soliton-induced Cherenkov radiation) when $k_{sol} = k_{lin}$ \cite{akhmediev1995cherenkov} resulting in
	\begin{equation}
	\label{eq:cherenk_lle}
	\dfrac{A^{2}}{2}=\dfrac{-2D_{int}}{\Delta{\omega}}-\alpha ,
	\end{equation}
	where $D_{int}$, shown in Fig. \ref{fig2}(a), is defined as $\omega_{\mu}-\omega_{0}-D_{1}\mu=\sum_{n=2}\dfrac{D_{n}}{n!}{\mu^{n}}$. Considering soliton amplitude in Fig. \ref{fig2}(c) with normalized detuning $\alpha = 3$, soliton-induced Cherenkov radiation would satisfy Eq. (\ref{eq:cherenk_lle}) at $\lambda = 743$ nm. The resulting wavelength is in good agreement with the peak observed (at 741 nm) in Fig. \ref{fig2}(a). The generated Kerr-comb spans from $715$ nm to $1070$ nm ($-70$ dB window), which covers from NIR to visible wavelength. This is, to the best of our knowledge, the first practical demonstration of Kerr-comb generation in thin SiN in the NIR to visible spectral range. This important property is achieved by engineering the waveguiding structure that is used to form the microring resonator.
	
\begin{figure*}
		\centering
		\includegraphics[trim=0cm 0cm 0cm 0cm,width=16cm,clip]{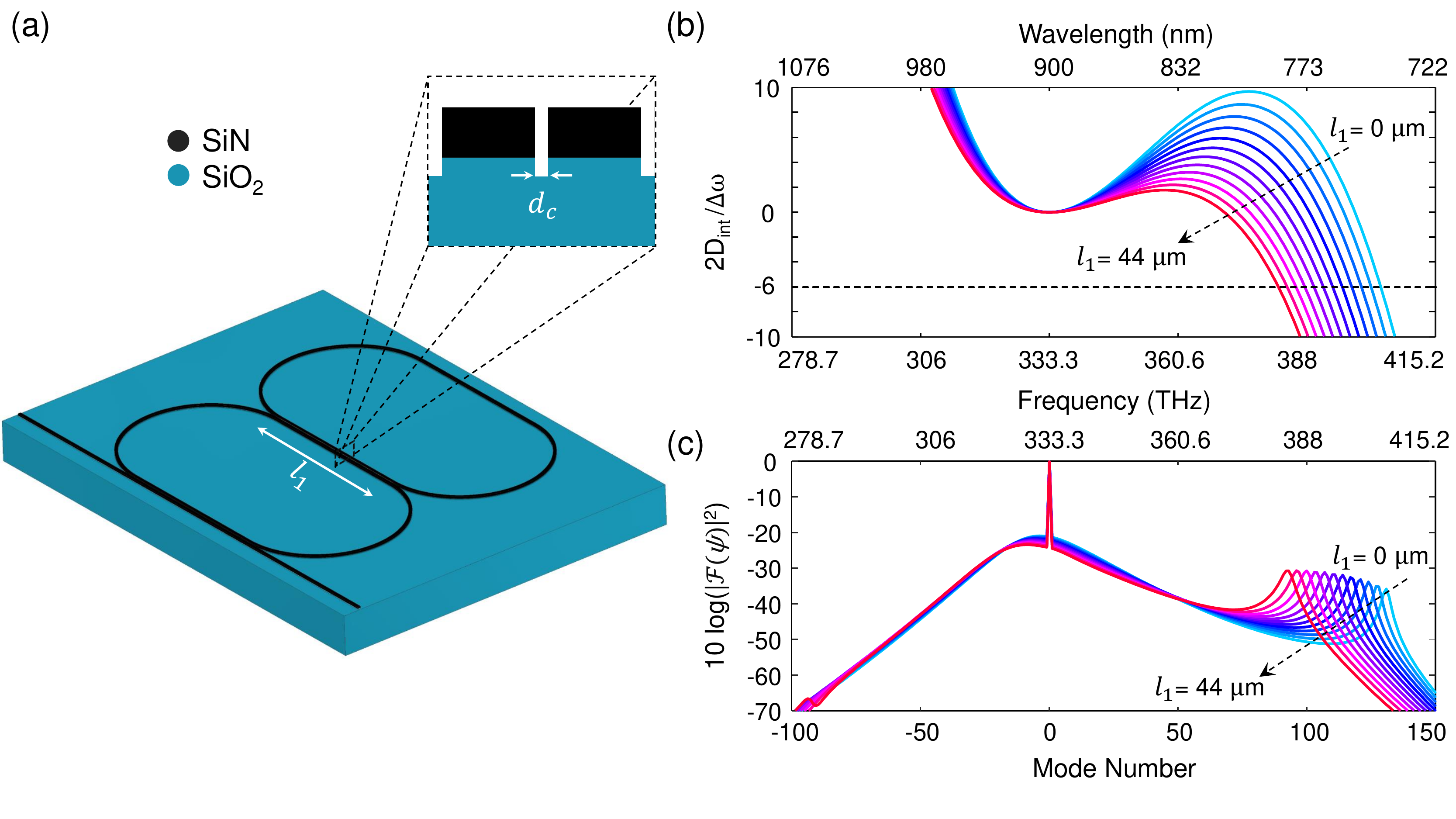}
		\captionsetup{justification=justified}
		\caption{Adjusting soliton Cherenkov radiation using a coupled-racetrack resonator: a) schematic of a coupled-resonator structure formed by coupling two identical racetrack resonators (based on the proposed waveguide) with the coupling gap of $d_c = 100$ nm and a coupling length ($l_1$) varying between $0$ and $44$ $\mu$m. b) Normalized integrated dispersion ($2D_{int}/{\Delta\omega}$) of the odd resonant eigenmode of the coupled-racetrack resonator in (a) for different effective coupling lengths ($l_1$) between resonators, ranging from zero (no coupling, i.e., single resonator case, far right) to 44 $\mu$m (far left) in steps of 4 $\mu$m. c) Generated Kerr-comb in the coupled-racetrack resonator structures in (a) for different coupling lengths ($l_1$) with normalized detuning of $\alpha = 3$ and normalized power of ${F}^2 = 3$. }\label{fig3}
	\end{figure*}

	For several applications, it is highly desired to control (e.g., tune) the Cherenkov radiation peak wavelength. Among the three main factors (i.e., pumping power, pumping wavelength detuning, and integrated dispersion), engineering the integrated dispersion (${D_{int}}$) is more desired for achieving a higher degree of control (e.g., wider tuning), as two other parameters are mainly adjusted to achieve a solitonic state. $D_{int}$ can be adjusted by either changing the dispersion of the original waveguide (that is used to form the resonator) \cite{oh2017coherent,johnson2015octave} or by engineering the mode of the resonator (without changing the original waveguide). The latter can be readily achieved by using a coupled resonator structure (using the same waveguide) in which the overall resonant modes (and their dispersion) are highly controlled by the coupling between the resonators. In contrast to previous reports of Cherenkov radiation manipulation based on changing the original waveguide profile and material \cite{oh2017coherent,johnson2015octave}, we primarily focus on resonant mode engineering in a coupled resonator structure to engineer the spectral property of Cherenkov radiation. We believe this approach provides a more practical way of using Cherenkov radiation for comb generation (among other applications) by enabling a wider range for controlling D$_{int}$ at higher speeds and lower powers.
	Figure \ref{fig3}(a) shows the schematic of two coupled racetrack resonators that are coupled to a bus waveguide for pumping. The use of racetrack resonators enables a simple way of controlling the coupling between the resonators through changing the flat waveguide regions (with length $l_1$) in Fig. \ref{fig3}(a).
	The coupling of the two identical resonators results in the splitting of each degenerated resonant frequency pair into two new (even and odd) eigenmodes with different (i.e., even or odd) electric field profile symmetries in the two resonators. The shift between the resonance frequencies of the two split modes is proportional to the coupling strength between the two resonators. Consequently, by engineering the coupling between the two resonators at different frequencies (i.e., frequency-dependent coupling), the dispersion of each set of resonant modes (even or odd modes) can be readily controlled (see Supplementary Information Section 1).
	
	By controlling the coupling between the two racetrack resonators in Fig. \ref{fig3}(a), we can adjust the soliton-induced Cherenkov radiation wavelength. The integrated dispersion of the odd resonant eigenmode of the coupled structure and its associated Kerr-comb spectrum with normalized detuning of $\alpha=3$, and normalized power of $F^{2}=3$ with a fixed coupling gap ($d_c = 100$ nm) in Fig. \ref{fig3}(a) and ranging coupling length ($0<l_1<44$ ${\mu}m$) are shown in Figs. \ref{fig3}(b) and $3$(c), respectively. Note that while $\alpha$ is the same as that used for the simulation of the single-resonator structure in Fig. \ref{fig2}, the actual pumping power here is twice that used in the single-resonator case because of the larger resonator length. Considering the numerically calculated soliton amplitude, soliton-induced Cherenkov radiation is expected to happen when the normalized integrated dispersion ($2D_{int}/{\Delta\omega}$) is equal to $-6$; which is in good agreement with the radiation peaks represented in Fig. \ref{fig3}(c). Figure \ref{fig3} clearly demonstrates the possibility of controlling the Cherenkov radiation wavelength over a reasonably large spectral range by simply changing the coupling length ($l_1$) without modifying the original waveguide structure or its material.\\
	
	Another advantage of the coupled resonator structure in Fig. \ref{fig3} over the single-resonator structure in Fig. \ref{fig2} is the $6$ dB increase in the Kerr-comb amplitude at Cherenkov radiation wavelength (from $-36$ dB to $-30$ dB as seen by comparing the corresponding peaks in Figs. \ref{fig2}(a) and $3$(c)). We attribute this to the increased overlap between the fundamental soliton tail and the Cherenkov radiation \cite{akhmediev1995cherenkov} in the coupled-resonator structure. A minor contributor to this effect can also be the red shift of the phase-matching wavelength between the fundamental soliton and the Cherenkov radiation, which increases the available power of the fundamental soliton at resonance wavelength (i.e., Cherenkov radiation), and thus results in a higher Cherenkov radiation amplitude. Note that an important precaution in increasing the Cherenkov radiation amplitude is the possibility of generating multiple solitons. If all these solitons are phase-matched to the Cherenkov radiations at its peak, they can reduce the bandwidth of the Cherenkov radiation (and thus the bandwidth of the generated Kerr-comb) due to phase mismatch at wavelengths away from the peak. A multi-solitonic state can be avoided by using phase-modulated pumping, which is shown to be able to generate a single-soliton state due to the introduced force on soliton \cite{taheri2015soliton}.
	
	Generation and spectral distribution of a bright soliton (e.g., for generation of a wideband Kerr-comb) highly depends on dispersion (i.e., anomalous dispersion is required). Extending the spectral width of the anomalous dispersion will extend the soliton amplitude in a larger bandwidth. In this way, if the phase-matching wavelength (which corresponds to the Cherenkov radiation) is adjacent to the anomalous dispersion, the radiation amplitude will be enhanced via stronger spectral overlap between the soliton and the Cherenkov radiation.
	
	One approach to generate anomalous dispersion in the coupled resonator structure is to modulate the group velocity dispersion of the resonator using an asymmetric Mach-Zehnder Interferometer (MZI) to provide an oscillatory effective coupling between the two resonators (see Supplementary Information Section 1). This way, anomalous dispersion in spectral regions can be provided in the coupled configuration where the single resonator has normal dispersion. Another mechanism, which may attribute to a higher Cherenkov radiation amplitude, is soliton spectral tunneling that transfers energy from an anomalous dispersion spectral region to another anomalous dispersion region through a normal dispersion region sandwiched between them acting as a barrier \cite{kibler2007soliton,guo2013understanding}. The newly generated anomalous dispersion spectral regions increase the probability of transferring photons in cascaded FWM to the Cherenkov radiation wavelength through the tunneling effect \cite{kibler2007soliton}. This approach can be applied to increase the dispersive wave amplitude and enhance the Kerr-comb power transfer toward the visible spectrum. Thus, one approach to increase the power transfer to shorter wavelengths through Cherenkov radiation is the modulation of the dispersion to both: $1$) enhance the soliton spectral power at the Cherenkov radiation wavelength, and $2$) introduce and enhance the tunneling effect. This is achieved in this work by designing a frequency dependent coupling between the two resonators in a coupled racetrack architecture, where the coupling has an oscillating characteristic versus the frequency.
	
\begin{figure*} 
		\centering
		\includegraphics[trim=0cm 0cm 0cm 0cm,width=16cm,clip]{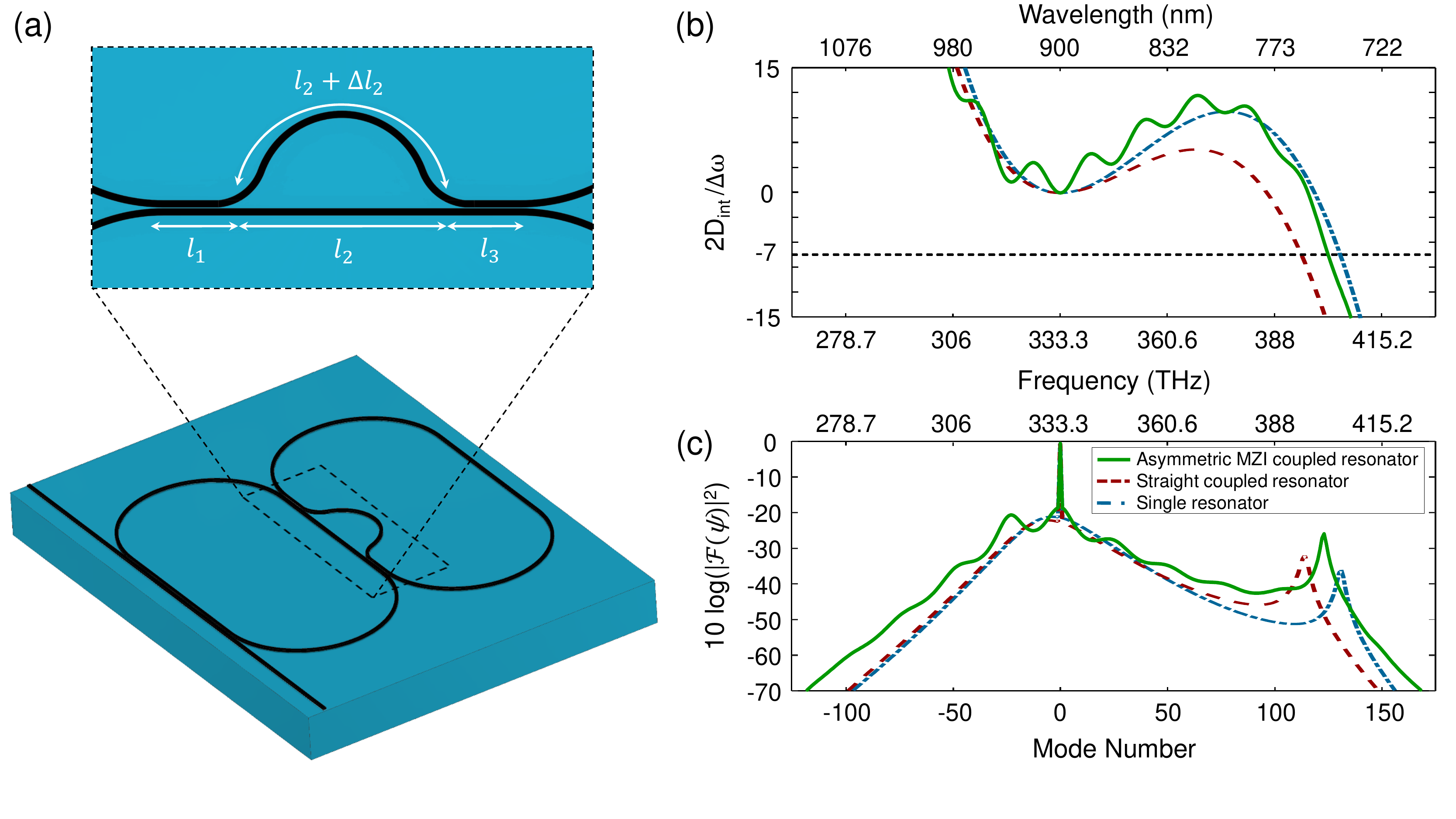}\\
		\captionsetup{justification=justified}
		\caption{Coupled racetrack resonators to increase both the dispersive wave amplitude, and the bandwidth of the generated Kerr-comb: a) schematic of a three-segment coupled racetrack resonator with coupling lengths ${l}_1$, ${l}_3$, and phase adjustment segment with length ${\Delta}{l}_2$ in one arm. b) Normalized integrated dispersion ($2D_{int}/{\Delta\omega}$), and c) Kerr-comb spectrum of the three-segment coupled resonator (solid green curve, with lengths $l_1 = 19$ ${\mu}m, l_2 = 1$ ${\mu}m, l_3 = 1$ ${\mu}$m, and ${\Delta}l_2 = 10$ ${\mu}m$), one segment coupled (dashed red, with coupling length $l_1 = 20$ ${\mu}m$), and single resonator (Fig. \ref{fig2}(a), dashed-dotted blue). Total length of all resonators are identical (i.e., $2\pi.40$ $\mu$m). Peak at Cherenkov radiation is increased for the coupled resonator, due to higher power density of fundamental soliton at resonance wavelength, when compared with a single resonator.}\label{fig4}
	\end{figure*}

	Figure \ref{fig4}(a) represents the schematic of a coupled-racetrack resonator, in which the coupling section is in the form of an asymmetric MZI composed of two separate coupling sections (with lengths $l_1$ and $l_3$) separated with a phase adjusting segment (with length $\Delta{l_2}$). The effective coupling between the two resonators leads to a sinusoidal perturbation on the overall resonator dispersion owing to the spectrally varying phase difference between the two coupling sections. The generated modulation in the resonator dispersion leads to consecutive spectral variation of dispersion between normal and anomalous dispersions, even in the frequency range where a single resonator has only normal dispersion. 
	
	Figure \ref{fig4}(b) depicts the normalized integrated dispersion of the odd eigenmode of a coupled racetrack resonator (green solid line) with the engineered three-segment MZI coupling representing the extension of anomalous dispersion regions toward Cherenkov radiation. The normalized integrated dispersion of the single resonator (dashed red line), and coupled racetrack resonators with straight coupling (dashed-dotted blue line) are depicted to represent effect of coupled resonators on the integrated dispersion. The Kerr-comb spectrum of the coupled-resonator structure in Fig. \ref{fig4}(a) is shown in Fig. \ref{fig4}(c), in which the normalized detuning $\alpha=3$, normalized power $F^{2}=3$ , coupling lengths $l_{1} = 19$ $\mu{m}$, $l_{2} = 1$ $\mu{m}$, $\Delta{l_{2}} = 10$ $\mu{m}$, and $l_{3} = 1$ $\mu{m}$ are assumed. Kerr-comb spectrum of coupled racetracks with straight coupling (i.e., $l_{1} = 20$ $\mu{m}$), and single ring resonators are shown in Fig. \ref{fig4}(c) for comparison, where all having the same total length ($2\pi. 40$ $\mu{m}$).
	
	Figure \ref{fig4}(c) shows that using the MZI-enhanced coupled racetrack resonators not only can increase the amplitude of the Cherenkov radiation but also can enhance the bandwidth of the comb spectrum: the bandwidth at $-70$ dB window is increased to $0.77$ of an octave and that at $-100$ dB reaches an octave spanning from $643$ nm to $1279$ nm.
	
\begin{figure*} 
		\centering
		\includegraphics[trim=7.5cm 0cm 0cm 0cm,width=16cm,clip]{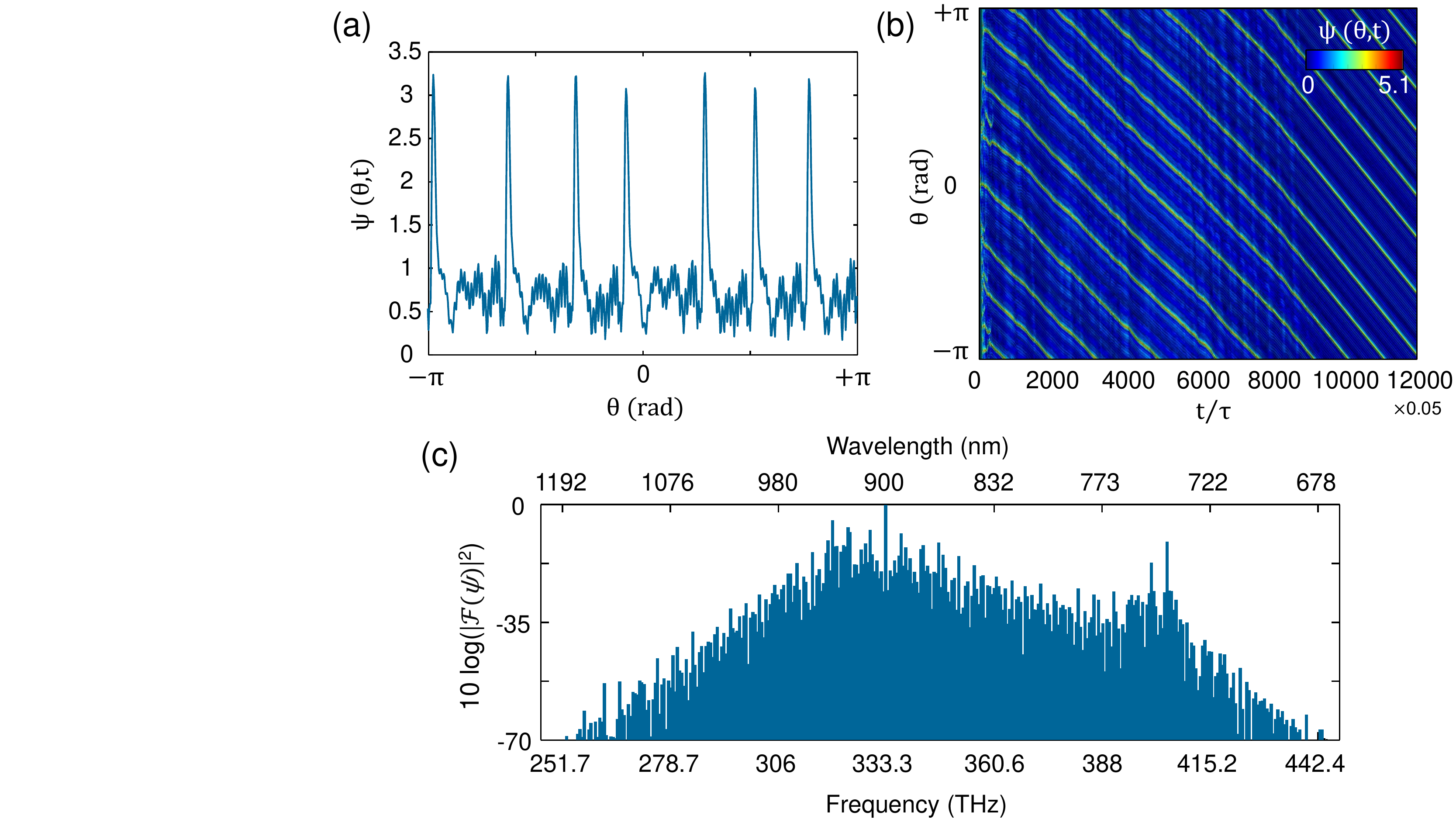}\\
		\captionsetup{justification=justified}
		\caption{Increasing peak associated with Cherenkov radiation in a multi-soliton state inside the coupled resonator structure in Fig. \ref{fig4}(a) with coupling lengths ${l}_1 = 18$ ${\mu}m$, ${l}_2 = 1$ ${\mu}m$, ${l}_3 = 1$ ${\mu}m$, phase adjusting segment with length ${\Delta}{l}_2 = 11$ ${\mu}m$, normalized detuning $\alpha = 3$ and normalized input power ${F}^2 = 5$: a) amplitude of the generated solitons. b) Time evolution of solitons in the multi-soliton state. c) Generated Kerr-comb in the coupled racetrack resonator.}\label{fig5}
	\end{figure*}

	By relaxing the single-soliton requirement, we have optimized the coupled racetrack structure to achieve the maximum power transfer to Cherenkov radiation at shorter wavelengths. In this optimization procedure, we have used an exhaustive search for coupling lengths (i.e., $l_1$ and $l_3$) and phase adjustment length (i.e., $l_2$) to find the maximum Cherenkov radiation peak while the normalized detuning is $\alpha = 3$, but the normalized power is increased to study both: 1) possibility of increasing bandwidth of the Kerr-comb associated to the solitonic states at $-70$ dB window, and 2) increasing Cherenkov radiation. Figure \ref{fig5} shows the optimized condition in which comb signals with enhanced Cherenkov radiation peak up to about $-10$ dB are achieved (here the structure in Fig. \ref{fig4}(a) with $l_{1}=18\ \mu$m, $l_{2}=1\ \mu$m, $l_{3}=1\ \mu$m, and phase adjusting segment with length $\Delta{l_{2}}=11\ \mu$m is considered along with $\alpha=3$ and $F^{2}=5$). Figure \ref{fig5}(a) shows the amplitude of solitons formed in the coupled microresonators, and Fig. \ref{fig5}(b) represents the temporal evolution of the generated solitons in the multi-solitonic state inside the coupled microresonators. Both Figs. \ref{fig5}(a) and $5$(b) clearly demonstrate the presence of multiple solitons in contrast to the previous cases (i.e., single resonator in Fig. \ref{fig2}(a) and coupled resonators in Fig. \ref{fig3}(a)) with a single soliton (see Figs. \ref{fig2}(a) and $3$(c), respectively). The comb spectrum shown in Fig. \ref{fig5}(c) clearly demonstrates two important properties: 1) considerably higher Kerr-comb peak compared to previous cases (Figs. \ref{fig2}(a) and $3$(c)), and 2) considerably higher Kerr-comb bandwidth (at any given minimum amplitude) compared to previous cases (Figs. \ref{fig2}(a) and $3$(c)). These two advantages are due to not only the increased pumping power, but also due to the engineered dispersion that makes power transfer more efficient to the Cherenkov radiation. The generated frequency comb spectrum in Fig. \ref{fig5}(c) clearly represents a modulation in spectral intensity which attributes to the multi-soliton state, where intensity of the resulting Kerr-comb can be approximated by $I=I_{s}\sum_{\eta}\mathrm{exp}(i(\theta-\theta_{\eta})\eta)$, where $\eta$ is the odd-supermode number and $\theta_{\eta}$ is the relative phase of the $\eta^{th}$ soliton. Although this configuration results a multi soliton state, phase modulation of pumping in the bus waveguide can be used to generate single soliton \cite{taheri2015soliton}.
	
	During the numerical analysis, we considered an input pump power of 451.8 mW in a single resonator and 0.9036 W (1.506 W) in the coupled-resonator structure corresponding to $F^2 = 3(5)$. This power can be reduced by decreasing $F$ and/or increasing the Q of the resonator. However, decreasing power results in reduction in the spectral peak amplitude at the Cherenkov radiation wavelength, and thus making the soliton spectrum narrower. This can be interpreted as a weaker soliton spectral component at the resonance wavelength. Considering -70 dB window for the Kerr-comb, spectral extension from NIR to the visible spectrum can be achieved by 150.42 mW for pump power in a single resonator and 300.8 mW in a coupled resonator structure for $F^2 = 2.1$ and $Q = 1.45 \times 10^5$. This input power results in -53 dB spectral peak at the resonance wavelength in a single resonator and -35.3 dB and -27 dB in coupled-resonator structures with one-segment coupling and asymmetric MZI coupling, respectively. This power requirement is easily achievable in practical integrated photonic structures. Thus, our MZI-based coupled-resonator structure can enable wideband Kerr-comb in visible-NIR wavelengths. This is to the best of our knowledge, the first presentation of an integrated photonics structure for visible Kerr-comb generation in a thin SiN waveguide with such a wide bandwidth operating at reasonable powers.
	\section{Conclusion}
	In conclusion, we demonstrated here an over-etched SiN waveguide structure to extend the anomalous group velocity dispersion into NIR wavelengths. Using this waveguide configuration, we showed Cherenkov radiation in the visible spectrum, which manifests itself through a peak in the generated Kerr-comb in the corresponding wavelength. By forming a coupled resonator structure using this waveguide configuration, this wavelength can be adjusted in a wide spectral range without changing pumping power, detuning or the waveguide configuration. We showed that coupled-racetrack resonators provide another tool for dispersion engineering through manipulating the eigenfrequencies of the resonant modes in the structure. Finally, by engineering the coupling spectrum of the coupled-resonator structure, we demonstrated enhancing the Cherenkov radiation up to $-10$ dB in a multi-soliton state, which results in wideband Kerr-comb generation in the visible-NIR wavelengths.
	\section{Acknowledgment}
	We would like to thank Razi Dehghannasiri, Amir Hossein Hosseinnia from Georgia Institute of Technology for helpful discussion. This research was supported by the Air Force Office of Scientific Research under Grant FA9550-15-1-0342 (Dr. Gernot Pomrenke)
	
	\newpage
	
	\newpage
	\newcommand{\beginsupplement}{%
		\setcounter{table}{0}
		\renewcommand{\thetable}{S\arabic{table}}%
		\setcounter{figure}{0}
		\renewcommand{\thefigure}{S\arabic{figure}}%
		\setcounter{section}{0}
		\renewcommand{\thesection}{S\arabic{section}}%
		\setcounter{equation}{0}
		\renewcommand{\theequation}{S\arabic{equation}}%
	}
	\beginsupplement
	\section{Supporting information}
		\subsection{Theoretical Framework}
		\label{sec:supplement theoretical}
		In order to calculate eigenfrequencies of two coupled identical resonators, we calculate equivalent coupling between the two resonators. To do so, transfer matrices for each part of the coupled configuration are written to complete a round-trip At first, the optical wave experiences a coupling whose transfer matrix is
		\begin{equation}
		\label{eq:supp1}
		\begin{pmatrix} t_m & jk_m\\
		j{k_m}^* & {t_m}^*
		\end{pmatrix} = \mathrm{exp}(\begin{pmatrix}
		-j\beta_1 & -jK_m\\
		-jK_m & -j\beta_2
		\end{pmatrix}l_m), 
		\end{equation}
		
		where $K_m$ is coupling coefficient, $\beta_{(1,2)}$ are propagation constants of the coupled waveguides on the two sides of the coupled region; $t_m$ and $k_m$ are transmission and coupling constants; $l_m$ is the coupling length between two coupled waveguides.
		After any coupling region, there could be a phase change section through which waves at one of the arms experience more delay due to the longer propagation length. Considering last part as a coupled section, the overall coupling which relates the input electric-field components (a$_i$, b$_i$) to the output electric-field components (a$_f$, b$_f$) in the coupling section is
		\begin{equation}
		\label{eq:supp2}
		\begin{pmatrix} a_f\\
		b_f
		\end{pmatrix} = 
		\begin{pmatrix} t_f & jk_f\\
		j{k_f}^* & {t_f}^*
		\end{pmatrix}\\
		\prod
		\begin{pmatrix}
		{\mathrm{e}}^{-j\beta_1(l_m+{\delta}l_m)} & 0\\
		0 & {\mathrm{e}}^{-j\beta_1(l_m)}
		\end{pmatrix}
		\begin{pmatrix} t_f & jk_f\\
		j{k_f}^* & {t_f}^*
		\end{pmatrix}
		\begin{pmatrix} a_i\\
		b_i
		\end{pmatrix}.
		\end{equation}
		In Eq. \ref{eq:supp2}, all parameters are the same as those named before, and $\delta l_m$ is the length of the phase adjustment section in the asymmetric MZI. 
		After the coupling section, another phase delay is experienced to complete a round-trip in each resonator. The resonators are considered identical; therefore, the overall round-trip length ($l_t$) would be the same for both resonators. The transfer matrix in a round-trip is
		\begin{subequations}
			\begin{equation}
			\begin{split}
			\begin{pmatrix} a_i\\
			b_i
			\end{pmatrix} = 
			\begin{pmatrix}
			{\mathrm{e}}^{-j\beta_1(l_{r1})} & 0\\
			0 & {\mathrm{e}}^{-j\beta_2(l_{r2})}
			\end{pmatrix}
			\begin{pmatrix} a_f\\
			b_f
			\end{pmatrix},\\
			\end{split}
			\end{equation}
			with
			\begin{equation}
			l_{r1} = l_t-\sum{l_m+\delta{l_m}},
			\end{equation}
			\begin{equation}
			l_{r2} = l_t-\sum{l_m}
			\end{equation}
		\end{subequations}
		Although the derived coupling matrix corresponds to the coupling of two resonators with an engineered coupling section (i.e., straight coupling or asymmetric MZI), we can assign an equivalent point coupling matrix by extracting phase delay experienced in each arm of the coupling section by each waveguide forming the coupled region. The effective point coupling matrix is
		\begin{equation}
		\begin{split}
		\label{eq:supp4}
		\begin{pmatrix} t & jk\\
		j{\tilde{k}} & \tilde{t}
		\end{pmatrix} = 
		\begin{pmatrix}
		\mathrm{e}^{j{\beta_1}l_t} &0\\
		0&\mathrm{e}^{j{\beta_2}l_t}
		\end{pmatrix}
		\begin{pmatrix}
		\mathrm{e}^{{-j}{\beta_1}l_{r1}} &0\\
		0&\mathrm{e}^{{-j}{\beta_2}l_{r2}}
		\end{pmatrix}
		\begin{pmatrix} t_f & jk_f\\
		j{k_f}^* & {t_f}^*
		\end{pmatrix}\\
		\prod
		\begin{pmatrix}
		\mathrm{e}^{{-j}{\beta_1}({l_m+{\delta{l_m}})}} &0\\
		0&\mathrm{e}^{{-j}{\beta_2}l_m}
		\end{pmatrix} 
		\begin{pmatrix} t_m & jk_m\\
		j{k_m}^* & {t_m}^*
		\end{pmatrix}
		\end{split}
		\end{equation}
		Using Eq. \ref{eq:supp4} for the equivalent coupling matrix between two coupled resonators with the same equivalent length, the approximate eigenfrequencies associated with the coupled resonators are 
		\begin{equation}
		\tilde{\omega}={\omega_0}\pm{\frac{c}{nl_t}}{\mathrm{cos}}^{-1}(\frac{t+\tilde{t}}{2}),
		\end{equation}
		where $\omega_{0}$ is the cold cavity angular frequency.
		This equation results in two sets of eigenfrequencies corresponding to even and odd eigenmodes in the coupled-resonator structure.
		\subsection{Numerical Simulation}
		Mode analysis of our proposed waveguiding configuration, (either single-waveguide or the coupled-waveguide structure) is performed using FEM in the COMSOL Multiphysics environment. The dispersion relation for SiN is found by fitting a single-pole sellmeier dispersion relation
		\begin{equation}
		\label{eq:supp6}
		\epsilon(\lambda) = \epsilon_{inf}+\frac{A{\lambda}^2}{{\lambda}^2-B^2}-E{\lambda}^2
		\end{equation}
		to our experimental data. In Eq. (\ref{eq:supp6}), $\epsilon_{inf} = 1$, $A = 3.005$, $B = 0.13624$, and $E = 0.01892$.
		\begin{tocentry}
			\includegraphics[trim=0cm 0cm 0cm 0cm,width=9cm,height=3.6cm,clip]{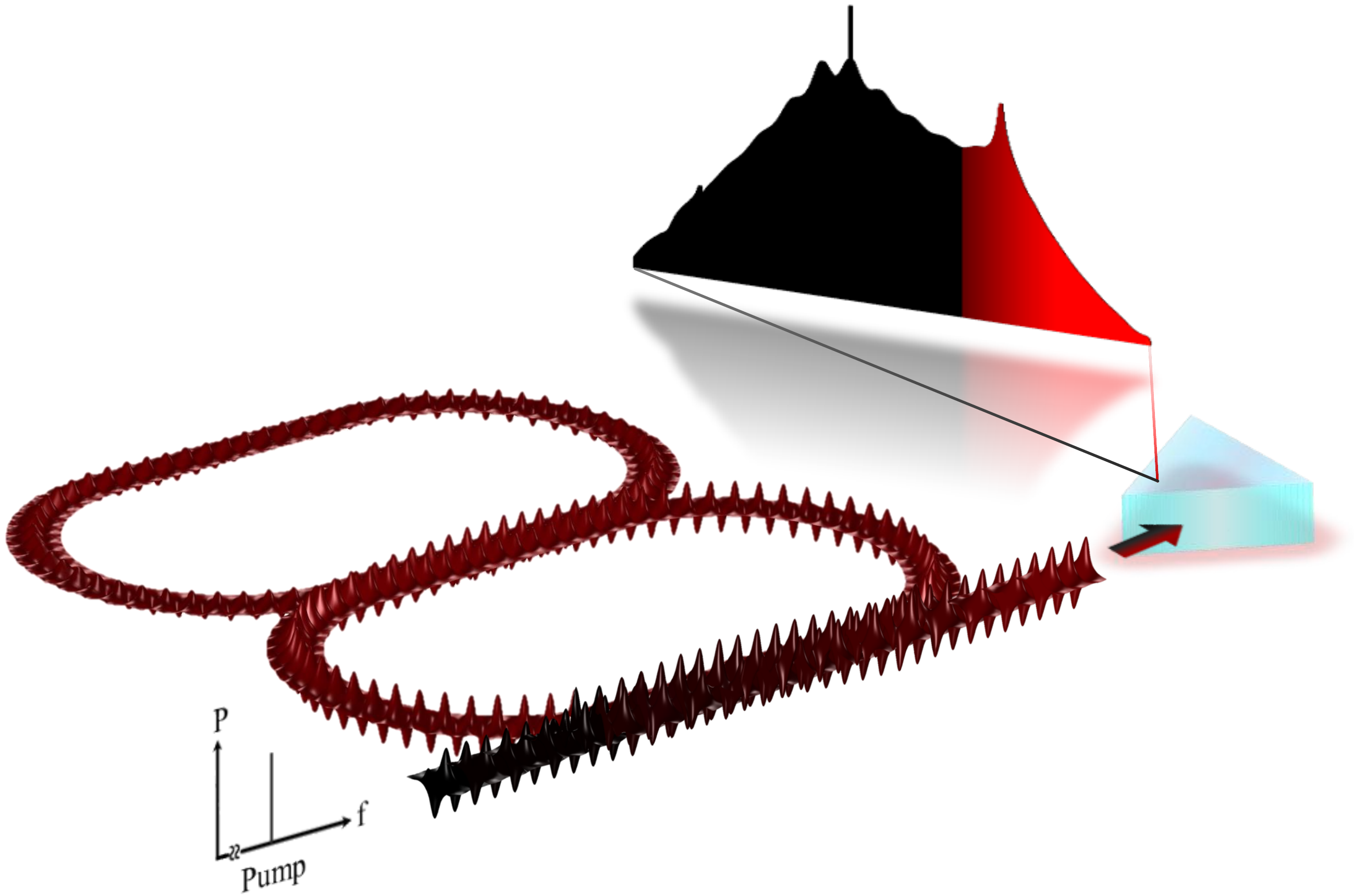}
		\end{tocentry}
	\newpage
	\providecommand{\latin}[1]{#1}
	\providecommand*\mcitethebibliography{\thebibliography}
	\csname @ifundefined\endcsname{endmcitethebibliography}
	{\let\endmcitethebibliography\endthebibliography}{}

\end{document}